\documentclass[%
prl,
 amsmath,amssymb,
 reprint,%
pr]{revtex4-1}

\usepackage{bm}
\usepackage[colorlinks]{hyperref}
\usepackage {graphicx}
\usepackage {amsmath}
\usepackage {hyperref}
\usepackage{upgreek}
\usepackage{gensymb}
\usepackage{soul}
\usepackage[usenames, dvipsnames]{color}

\def\gev{GeV/$c^2$}

\begin{document}


\title{Crystal Defects: A Portal To Dark Matter Detection}


\author{Fedja Kadribasic, Nader Mirabolfathi}
\affiliation{Department of Physics and Astronomy, Texas A\&M University, College Station, TX, USA}

\author{Kai Nordlund and Flyura Djurabekova}
\affiliation{Helsinki Institute of Physics and Department of Physics, PB 43, University of Helsinki, Finland}


\date{\today}

\begin{abstract}
We propose to use the defect creation energy loss in commonly used high energy physics solid state detectors as a tool to statistically identify 
dark matter signal from background. We simulate the energy loss in the process of defect creation using density functional theory and molecular dynamics methods and calculate the corresponding expected dark matter spectra. We show that in phonon-mediated solid state detectors, the energy loss due to defect creation convolved with the expected dark matter interaction signal results in a significant change in the expected spectra for common detector materials. With recent progress towards $\sim$10 eV threshold low-mass dark matter searches, this variation in expected dark matter spectrum can be used as a direct signature of dark matter interactions with atomic nuclei. 
\end{abstract}


\maketitle

A plethora of evidence indicates that around 85\% of the Universe's matter is nonbaryonic \cite {Ade:2013zuv}. The nature of this so called Dark Matter (DM) is yet to be deciphered. Although historically the focus of most of the dark matter direct search experiments has been on particles with the mass of a few tens of 10 \gev,  in recent years there has been a paradigm shift to include lower mass ($<$10 \gev) dark matter candidates \cite {Cushman:2013zza}. Due to ${m_{DM}}^2$ dependence of the expected DM-nucleus interaction cross section, very low energy threshold ($\sim$10 eV) detectors are required for their detection. Dual measurement techniques such as those used by CDMS, CRESST, and similar experiments fail at these low thresholds due to the secondary measurement fundamental noise and the quantum excitation limitations, which motivates research into alternative methods to discriminate nuclear recoils (signal) from the electron recoils (noise). In this work, we show how the energy required to produce a defect, which we refer to as the defect creation energy ``loss'', can be used to produce a signature for a potential dark matter signal. Major progress has been made in developing single-electron sensitive detectors \cite {Contactfree, Lee:2020abc, PD2_pres, Agnese:2018col, contactfree_2, contactfree_ltd, Abdelhameed:2019hmk} and next generation phonon-mediated detectors are reaching the sensitivities that are required to measure energy loss due to defect creation process in crystals.  

Penetrating high energy radiation of all forms produces long-lived defects in materials \cite {Averback97, Nor18}. This includes high-energy electromagnetic radiation, such as gamma rays, that primarily interact with the electrons in solid-state detectors, and other particles such as neutrons, which interact primarily with the nuclei \cite {Speller_2015, Mercure03, Nor18b}. At the interaction energies approaching the eV-scale, photons (and more generally electron recoil events) can transfer enough energy to electrons to create an electron-hole pair excitation but not enough to produce stable defects \cite {Romani:2017iwi, Hol08a, Hol09a, Hol10a}.  If the incident particle scatters off the nucleus, the energy required to produce a defect  \cite {Agnese:2018nbs} can be directly delivered to the nucleus and create long-lived defects. Additionally, the threshold displacement energy strongly depends on the direction of the recoiling atom and target material \cite {Nor18, Nor18b, Vaj77}. Hence, the fact that low-energy nuclear recoils produce long-lived defects whereas low-energy electron recoils do not means that, in principle, the energy loss due to defect creation can be used to discriminate nuclear recoils from electron recoils. In particular, we expect to see an apparent energy shift for those recoils whose energies are more than the defect creation threshold. In the case of dark matter elastic nucleus scattering, this change in the expected energy spectrum can be used as a tool to statistically discriminate backgrounds. 

We use molecular dynamics simulations with classical potentials backed by density functional theory (DFT) calculations to find the energy loss as a function of recoil energy and nuclear recoil angle. We simulate defect creation with the PARCAS code \cite {
Nor97f, Nor05c} 
for elemental carbon (C), silicon (Si), and germanium (Ge) atoms in a diamond cubic lattice structure. Each element is initialized as a unit cell of 4096 atoms that is thermalized to 40 mK, and the potential energy of the entire system is measured afterwards. A random atom from the central 64 atoms is selected and displaced in a random direction with an energy from 1 to 200 eV. Contrary to our previous work \cite {Kadribasic:2017obi} where the energy scan was stopped when the threshold in a given direction was reached, in the current work we always simulate up to 200 eV to get the stored energy for above-threshold recoils as well. After waiting for 10 ps, the amount of time necessary for a defect to stabilize \cite {Hol08a, Hol10a}, the energy of the system is calculated. The difference between the final energy of the system and the initial energy gives the defect creation energy loss since that is the energy that is not converted into phonons. We use the Stillinger-Weber potential \cite {Sti85, Din86} for Si and Ge, and we use Tersoff potential extended by Nordlund \cite {Ter89, Nor96} and Erhart \cite {Erh04} potentials for C. These potentials were chosen because they give good agreement with experimental \cite {Koi92, Bou76} or quantum mechanical density functional theory threshold energies \cite {Hol08a, Hol10a}. 347 recoil directions are simulated for the C-Erhart data, 343 for C-Tersoff-Nordlund, 6591 for Ge, and 1003 for Si. 

\begin{figure}
\begin{center} \includegraphics[width=0.98\columnwidth]{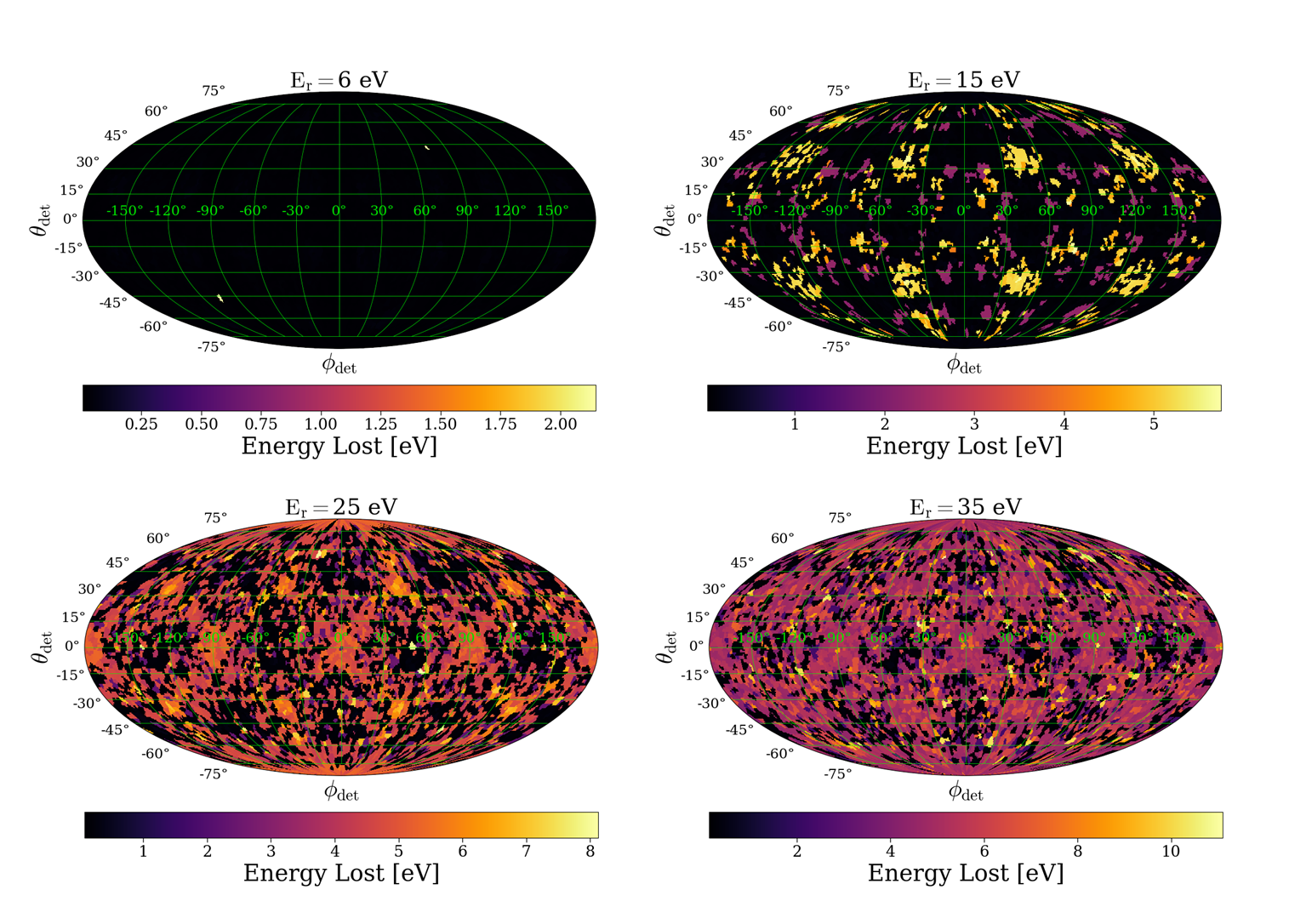} \end{center}
\caption{\label{Eloss}
Four Mollweide-projection plots showing the defect creation energy loss in germanium at recoil energies of 6 eV (upper left), 15 eV (upper right), 25 eV (lower left), and 35 eV (lower right). The color of the region on each plot indicates the amount of energy that went into creating a defect in a particular direction at a particular energy. Below about 6 eV in germanium, the recoil is not strong enough to create a defect, hence why most of the 6 eV recoil energy plot is dark. As the recoil energy increases, the defect creation energy loss is smeared out and approaches a linear regime at energies greater than about 100 eV in germanium. 
}
\end{figure}

\begin{figure}
\begin{center} \includegraphics[width=0.98\columnwidth]{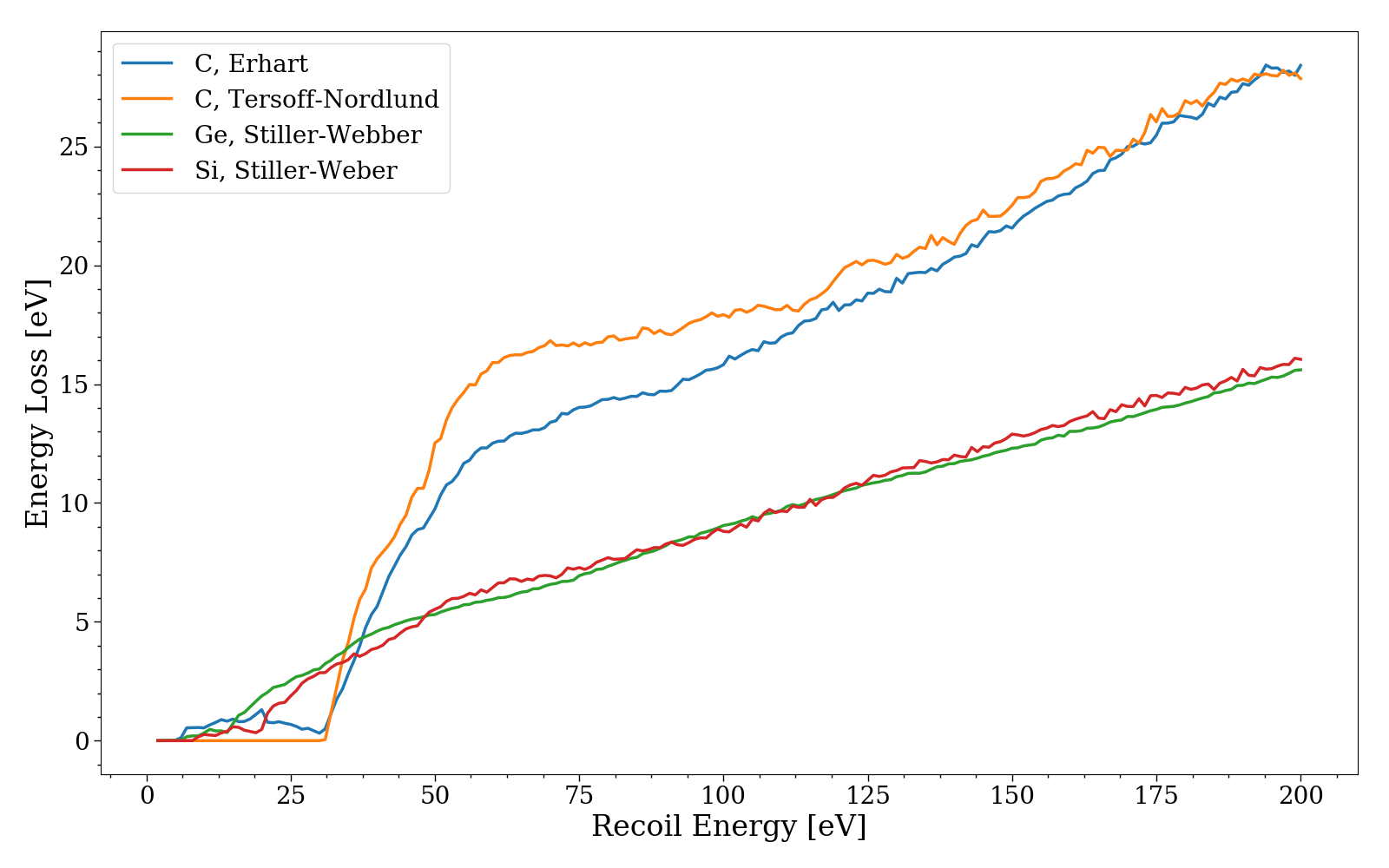} \end{center}
\caption{\label{Eloss_mean}
Defect creation energy loss over all recoil directions for several materials. We employ two models for carbon - Erhart (blue) and Tersoff-Nordlund (orange). Since it is difficult to display all of the data used in this study, this plot summarizes the functional form of the energy loss as a function of recoil energy. 
}
\end{figure}

A summary of the energy loss results thus obtained is shown in Fig. \ref {Eloss} and Fig. \ref {Eloss_mean}. Fig. \ref {Eloss} shows the energy loss as a function of recoil energy and recoil angle for several representative recoil energies. The fact that there is a periodic variation in energy loss as a function of recoil angle indicates that the energy loss is anisotropic, as expected from the crystal lattice structure. Additionally, the dark regions demonstrate that this is a stochastic process that does not always produce defects; however, when a defect is produced, the effect can be very pronounced, especially at recoil energies of a few tens of eV. Fig. \ref {Eloss_mean} shows the mean energy loss as a fuction of recoil energy 
for all three elements. 
The energy loss has the strongest effect at energies of a few tens of eV when the energy loss can be comparable to the energy of the recoiling nucleus, and the variation in energy loss is also consistently larger at these low energies than it is past about 100 eV. 

To find the effect of the energy loss on the dark matter spectrum, we calculate $\dfrac {\partial^2 R} {\partial E_r \partial \Omega_r}$ at the SNOLAB site (46.4719\degree, 81.1868\degree) on September 6, 2015 (to match \cite {rate}) for several dark matter masses. This is found via the integral outlined in \cite {Kadribasic:2017obi} to find the differential rate per unit recoil energy for a perfect detector. We choose $2 \times 10^8$ sample events from the distribution given by $\dfrac {\partial^2 R} {\partial E_r \partial \Omega_r}$ and add energies sampled from a gaussian distribution centered at 0 eV with standard deviation of 1 eV. The integral of this result gives the expected differential rate assuming the perfect detector has 1 eV recoil energy resolution. To find the effect of the energy loss, we use the sampled events from the previous step and subtract the numerically calculated energy loss for recoil energies 2-170 eV. For recoil energies greater than 170 eV, we calculate a linear functional fit for the energy loss using the mean of the angle-dependent energy loss data from 100 to 200 eV. 

\begin{figure}
\begin{center} \includegraphics[width=0.98\columnwidth]{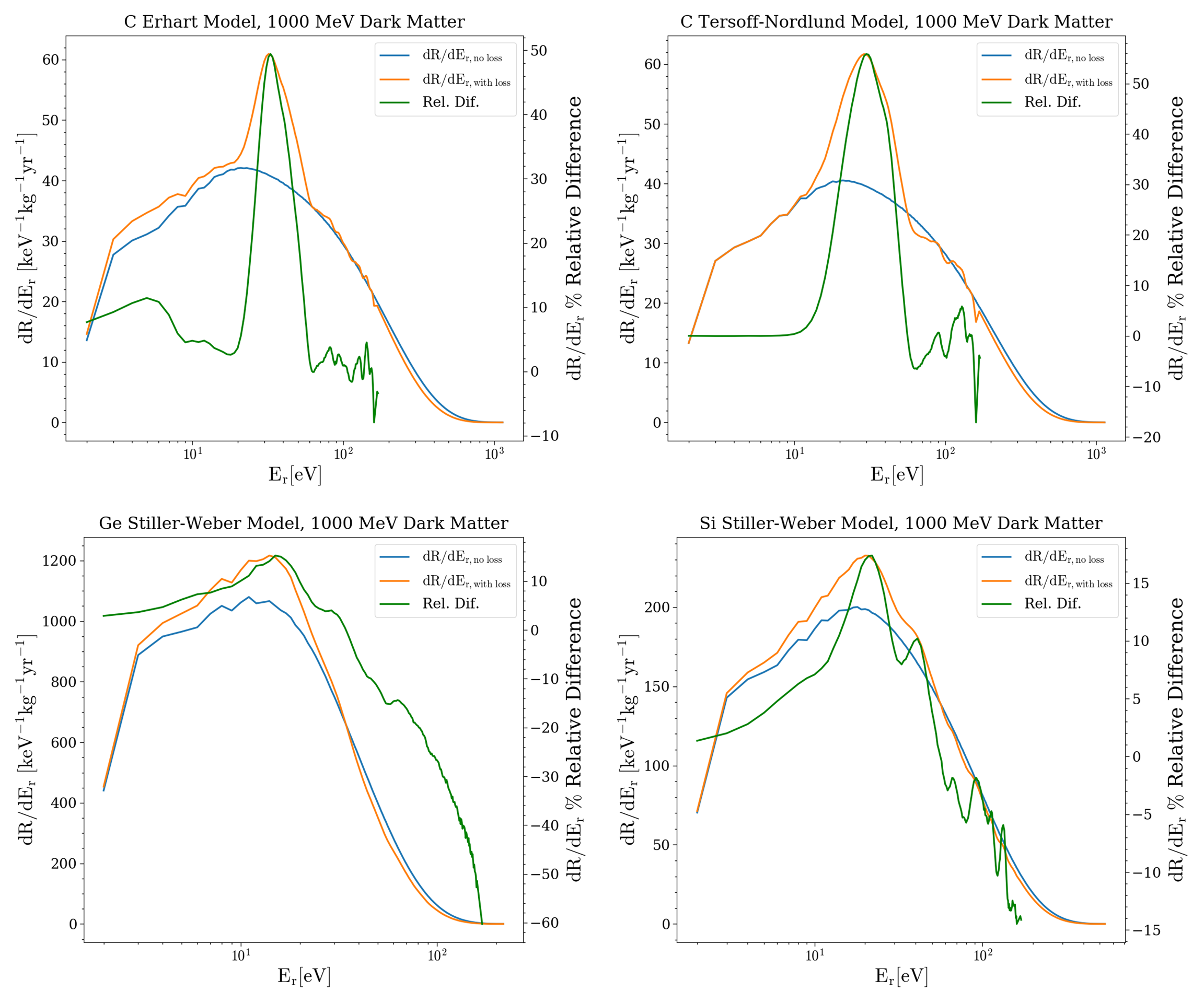} \end{center}
\caption{\label{dR_element_3_eV}
Differential rate per unit recoil energy as a function of recoil energy for the four models used in this work. The upper two plots are for carbon (Erhart model at left and Tersoff-Nordlund model at right), and the lower two plots are for germanium (left) and silicon (right). All four plots assume dark matter with particles of mass 1 GeV$/c^2$, calculated on September 6, 2015 at SNOLAB for $10^{-42} \text {cm}^2$ cross section dark matter. Blue curves show the effect of a 3 eV detector resolution with 10 eV threshold. Orange curves show the result of including the numerically-calculated energy loss to defect creation, i.e. energy that cannot be measured by the detector. Green curves show the relative difference between differential rate with and without the energy loss effect included up to 170 eV recoils. $2 \times 10^8$ events are used for these simulations. 
}
\end{figure}

The results of the above calculation are shown in Fig. \ref {dR_element_3_eV} for the four detector material models and 1 GeV$/c^2$ dark matter. Both figures show the differential rate per unit recoil energy as a function of recoil energy for a detector with a given resolution and threshold without (blue) the energy loss and with it (orange) taken into account. The kind of material used for the detector, as well as the model for the potential function in the lattice, has a significant effect on the size and location of the features that appear due to defect creation energy loss. Although the ultimate resolution of the detector has an effect on whether fine features due to the energy loss are resolved, the large features in the spectrum for carbon, irrespective of model, can still be resolved. Additionally, the location of the energy loss peaks relative to the high-energy recoil energy tail can, given a large enough signal, be used to determine the mass of the dark matter particle. 

\begin{figure}
\begin{center} \includegraphics[width=0.98\columnwidth]{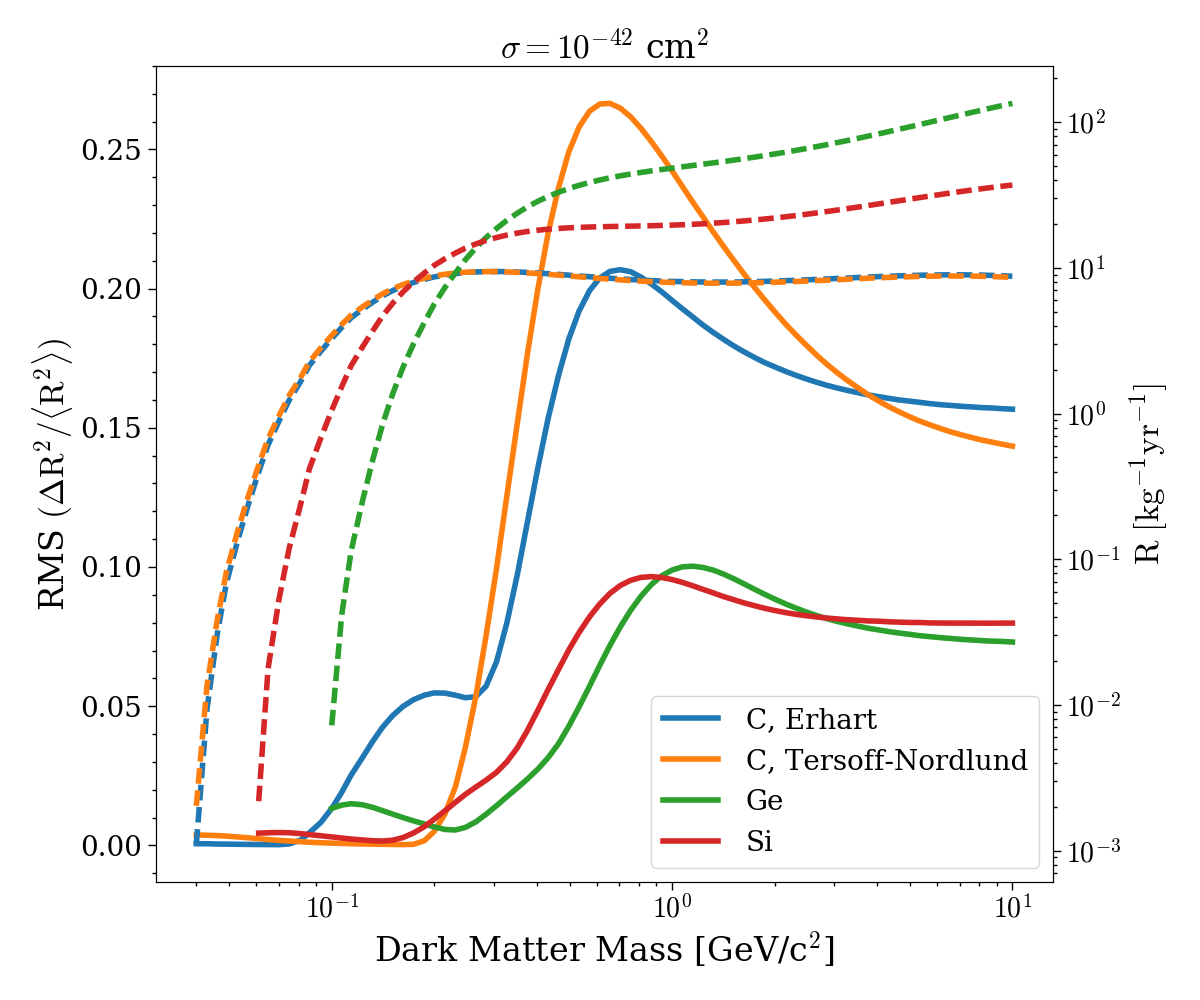} \end{center}
\caption{\label{RMS}
Signal strength due to defect creation energy loss, as normalized RMS, on the left y-axis and total integrated rate on the right y-axis, both as a function of dark matter mass. The normalized RMS quantity gives a qualitative measure of the ability of a potential dark matter spectrum to be differentiated from the noise floor given the features from defect creation energy loss. In this case, we assume that the dark matter spectrum without energy loss approximates the noise floor to show that a detector with sufficient resolution and a high enough dark matter rate can observe features in an, otherwise, featureless spectrum. The integrated rate is found by integrating the differential rate over all recoil energies given the energy loss effect. $2 \times 10^8$ events are used for these simulations, and a detector with 3 eV resolution and 10 eV threshold is assumed. 
}
\end{figure}

We further quantify the discrimination power by calculating a normalized root-mean-squared (normalized RMS) statistic as a function of dark matter mass. The procedure for doing so is the same as in \cite {Kadribasic:2017obi} with the exception that the fluctuations in the spectrum due to energy loss are compared to the case of a perfect detector with 3 eV resolution and 10 eV threshold. Mathematically, this is given by
\begin {equation}
\label {rmseq} \text {RMS}_{\text {norm}} = \sqrt {
\dfrac {\oint_{\Delta E} {\left(R_{\text {loss}} - R_{\text {no loss}}\right)^2 dE}}
{\oint_{\Delta E} {R_{\text {no loss}}^2	 dE}}
}
\end {equation}
In other words, the normalized RMS is found by looking at the squared difference between the orange and blue curves divided by the squared integral of the blue curves in Fig. \ref {dR_element_3_eV}. 

Fig. \ref {RMS} shows the result of doing so for all four models over a range of dark matter masses for a detector with 3 eV resolution and 10 eV detector threshold. The left y-axis indicates the normalized RMS statistic, whereas the right y-axis indicates the total rate found by integrating the differential rate, like that in Fig. \ref {dR_element_3_eV}, over all recoil energies given the defect creation energy loss effect. Although this analysis cannot determine exactly how effective this method is in resolving a dark matter signal from a given noise, it does give an idea of what range of dark matter masses it is most useful for and what detector materials show the strongest effect. Of the three materials investigated in this study, the signal strength would be by far the strongest in carbon, independent of model used, as long as the bulk of the events are not near the detector threshold. For this reason, in addition to the other useful properties of diamond detectors described in \cite {Kurinsky:2019pgb}, diamond detectors need to be developed to verify their theoretical capabilities for finding dark matter. 

In summary, we have described a method that could be used to measure a signature of dark matter detectors. Out of the three detector materials considered in this study, diamond detectors hold the most promise for making this vision a reality. These results, in conjunction with other recent studies such as \cite {Kurinsky:2019pgb}, corroborate the necessity to investigate novel detector materials. With many experiments coming online that aim to calibrate detectors for low energy nuclear recoils, such as the IMPACT@TUNL measurement \cite {TUNL_pres}, the way is paved not only for novel analysis tools for dark matter search experiments but also for understanding defect creation on an experimental rather than purely computational level. 

N.\ M.\ acknowledges Mitchell Institute For Fundamental Physics financial support. Kai Nordlund and Flyura Djurabekova acknowledge grants of computer capacity from the IT Centre for Science in Finland, CSC and the Finnish Grid and Cloud Infrastructure (persistent identifier urn:nbn:fi:research-infras-2016072533 ).

\bibliographystyle{apsrev4-1}
\bibliography{Bib_01}

\end{document}